\documentclass[11pt,letterpaper]{article}
\usepackage{graphicx,epsf,parskip,times,amsmath,lscape,multirow,amsfonts}
\usepackage[comma,sort&compress]{natbib}
\usepackage{caption}
\usepackage{array}
\newcolumntype{C}[1]{>{\centering\arraybackslash}p{#1}}
\newcommand{\lb}{\left(}
\newcommand{\rb}{\right)}

\newcommand{\lcb}{\left\{}
\newcommand{\rcb}{\right\}}

\newcommand{\lsb}{\left[}
\newcommand{\rsb}{\right]}

\newcommand{\indep}{\perp\!\!\!\perp}

\newcommand{\Tc}{T^{\underline{a}_0}}

\newcommand{\aUt}{\mathcal{A}_{U}}
\newcommand{\aLo}{\mathcal{A}_{L0}}
\newcommand{\aaLo}{\mathcal{A}_{L0,1}}
\newcommand{\aLl}{\mathcal{A}_{L1}}

\title{Simulating longitudinal data from marginal structural models using the additive hazard model}

\author{Ruth H. Keogh$^1$, Shaun R. Seaman$^2$, Jon Michael Gran$^3$, Stijn Vansteelandt$^{1,4}$}

\date{{\small$^1$Department of Medical Statistics, London School of Hygiene \& Tropical Medicine, Keppel Street, London, WC1E 7HT, UK\\
$^2$MRC Biostatistics Unit, University of Cambridge, Institute of Public Health, Forvie Site, Robinson Way, Cambridge CB2 0SR, UK\\
$^3$Oslo Centre for Biostatistics and Epidemiology, Department of Biostatistics, Institute of Basic Medical Sciences, University of Oslo, P.O. Box 1122 Blindern, 0317 Oslo, Norway\\
$^4$Department of Applied Mathematics, Computer Science and Statistics, Ghent University, 9000 Ghent, Belgium}}

\begin{document}
	
	\maketitle

\begin{abstract}
Observational longitudinal data on treatments and covariates are increasingly used to investigate treatment effects, but are often subject to time-dependent confounding. Marginal structural models (MSMs), estimated using inverse probability of treatment weighting or the g-formula, are popular for handling this problem. With increasing development of advanced causal inference methods, it is important to be able to assess their performance in different scenarios to guide their application. Simulation studies are a key tool for this, but their use to evaluate causal inference methods has been limited. This paper focuses on the use of simulations for evaluations involving MSMs in studies with a time-to-event outcome. In a simulation, it is important to be able to generate the data in such a way that the correct form of any models to be fitted to those data is known. However, this is not straightforward in the longitudinal setting because it is natural for data to be generated in a sequential conditional manner, whereas MSMs involve fitting marginal rather than conditional hazard models. We provide general results that enable the form of the correctly-specified MSM to be derived based on a conditional data generating procedure, and show how the results can be applied when the conditional hazard model is an Aalen additive hazard or Cox model. Using conditional additive hazard models is advantageous because they imply additive MSMs that can be fitted using standard software. We describe and illustrate a simulation algorithm. Our results will help researchers to effectively evaluate causal inference methods via simulation.
\end{abstract}

	\section{Introduction}
	\label{sec:intro}
	
	Observational longitudinal data are increasingly used to investigate the effects of treatments and exposures on health outcomes. To estimate treatment effects from observational data we must account for confounding of the treatment-outcome association, sometimes referred to as `confounding by indication', and recent years have seen huge developments in statistical and epidemiological methods for this task. In this paper we focus on the setting of estimating the joint effects of treatment across time-points on a time-to-event outcome using longitudinal data on treatment use and covariates, where time-dependent confounding is a specific challenge. When there is time-dependent confounding, standard analysis methods, such as Cox regression with adjustment for baseline or time-updated covariates, do not in general enable estimation of the causal effects of interest \citep{Daniel:2013}.  
	
	Several methods have been described for estimating the causal effects of longitudinal treatment regimes on time-to-event outcomes. Marginal structural models (MSM) estimated using inverse probability of treatment weighting (IPTW) for time-to-event outcomes were introduced by \cite{Hernan:2000}, who described use of marginal structural Cox models (Cox MSM). Other methods include estimation of MSMs using the g-formula (also called g-computation) \citep{Robins:1986,Daniel:2011,Keil:2014}, structural nested accelerated failure time models \citep{Robins:1992Bka,Hernan:2005}, structural nested failure time models \citep{Robins:1992Epi, Vansteelandt:2010}, structural nested cumulative failure time models \citep{Picciotto:2012}, and structural nested cumulative survival time models \citep{Seaman:2019}. A recent review \citep{Clare:2018} found that among these, the Cox MSM approach is by far the most commonly used method in practice. 
	
	With the increasing development of more advanced causal inference methods, it is important to be able to evaluate method performance in different scenarios and make comparisons between methods to guide their use in practice. Simulation studies are a key tool for such investigations and can be used to assess properties such as bias, efficiency and coverage of confidence intervals. The results help analysts to choose which methods are most appropriate for answering research questions using their data. The importance of well-conducted simulation studies was highlighted by \cite{Morris:2019}, who provide detailed guidance for their planning and reporting. In this paper we focus on the use of simulation studies for evaluations involving MSMs in the setting of a time-to-event outcome using longitudinal data on treatment use and covariates. When conducting a simulation study, it is desirable to be able to generate the data in such a way that the correct form of any analysis model to be fitted to those data is known, so that we know that the analysis model is correctly specified. For example, suppose that we wished to use a simulation study to assess the performance of the IPTW estimation approach for MSMs when the models for the weights are mis-specified in some way. It would be important to know that the MSM itself is correctly specified, so that any bias in the estimates can be attributed to mis-specification of the models used for the weights. As a second example, suppose that we wished to use a simulation study to compare the relative efficiency of the estimates of survival probabilities obtained using IPTW and using the g-formula. To make a fair comparison, the models involved in each approach should be correctly specified. 
	
	Generating longitudinal and time-to-event data in such a way that the form of models used in methods applied to the data are known is not straightforward. A reason for this is that it is natural for the data to be generated in a sequential conditional manner, generating each individual's covariates, treatment status, and survival status at each measurement time in turn conditional on the past, starting at time zero. This makes use of conditional models, including conditional hazard models for the time-to-event component. Analysis methods based on MSMs, on the other hand, make use of marginal (population average) rather than conditional hazard models. In this paper we show how to simulate longitudinal data on treatments and covariates together with a time-to-event outcome in such a way that the form of the MSM that specifies the marginal hazard of the outcome is known, and hence that we know or are able to derive the true values of its parameters and of causal estimands of interest such as risk differences or risk ratios. Our results will help researchers to effectively evaluate causal inference methods via simulation; a task of high importance but which is currently very rarely performed.
	
	We provide general results that enable the form of the correctly specified MSM to be derived from a conditional hazard model used in the data simulation procedure, and show how the results can be applied when the conditional hazard model is an additive hazard model \citep{Aalen:2008} or a Cox model \citep{Cox:1972}. We show that there is an advantage to using conditional additive hazard models for the data simulation, because this results in an additive form for the MSM, which can be fitted using standard software. The same does not hold for the Cox model. \cite{havercroft:2012,Young:2010} and \cite{Young:2014} outlined algorithms for simulating longitudinal and time-to-event data to correspond with a specified Cox MSM, but their methods require restrictive assumptions, which limits the simulation scenarios that can be generated. We instead place an emphasis on use of additive hazard models, and the scenarios to which our results can be used are not limited, as in the earlier work.
	
	The paper is organised as follows. In Section \ref{sec:setup} we outline the longitudinal data set up and the notation. In Section \ref{sec:msms} we review briefly why standard methods of analysis based on regression adjustment do not estimate the causal effects of interest and describe the use of MSMs in causal inference. Our main results are presented in Section \ref{sec:sim.msm}, where we derive formulae that describe the relationship between a conditional hazard model and an MSM for the hazard and show the advantages of simulating data using an additive hazard model. In Section 5 we provide an example simulation algorithm and the algorithm is illustrated in Section 6. R code corresponding to the algorithm and the illustration is provided at https://github.com/ruthkeogh/causal\_sim. We conclude with a discussion in Section \ref{sec:discussion}.
	
	\section{Longitudinal data and time-dependent confounding}
	\label{sec:setup}
	
\begin{figure}
		\caption{Causal directed acyclic graph (DAG) illustrating relationships between treatment $A$, time-dependent covariates $L$, an unmeasured frailty term $U$ and time-to-event, illustrated for a discrete-time setting where $Y_k=I(T>k)$.}\label{fig:dag}
		\centering
		\includegraphics[scale=1]{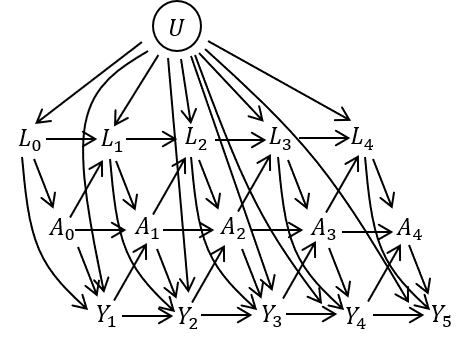}
	\end{figure}

	We consider a study in which $n$ individuals are observed at regular visits up until the earlier of the time of the event of interest and the censoring time. The visit times, assumed to be the same for everybody, are $k=0,1,\dots,K$. At each visit we observe binary treatment status $A_{k}$ and a set of time-dependent covariates $L_{k}$. A bar over a time-dependent variable indicates the history, that is $\bar{A}_{k}=\{A_0,A_1,\ldots,A_k\}$ and $\bar{L}_{k}=\{L_0,L_1,\ldots,L_k\}$. We let $\underline{A}_{k}=\{A_k,A_{k+1},\ldots,A_K\}$ denote treatment from visit $k$ up to $K$. The event time is denoted $T$. For simplicity we assume that all censoring is administrative at time $K+1$, but the analysis methods that we focus on in this paper also accommodate loss to-follow-up and we discuss this in Section \ref{sec:discussion}. Temporal causal relationships between variables are illustrated using a directed acyclic graph (DAG) in Figure \ref{fig:dag}. The DAG also includes a variable $U$, which has direct effects on $L_k$ and $T$ but not on $A_k$. $U$ is an unmeasured individual frailty and we include it because it is realistic that such individual frailty effects exist in practice. Because $U$ is not a confounder of the assocation between $A_k$ and $T$, the fact that it is unmeasured does not affect our ability to estimate causal effects of treatments. In the DAG the relationships are illustrated for a discrete-time setting where $Y_k=I(T>k)$. One can imagine extending the DAG by adding a series of small time intervals between each visit, at which $I(T>t)$ is observed. As the time intervals become very small we approach the continuous time setting. 
	
	It is possible to use the longitudinal data to estimate the impact of treatment at visit $k$, $A_k$ on the concurrent hazard, for example using a Cox regression with time-updated treatment and with adjustment for confounding by the past treatment and covariate history, $(\bar{A}_{k-1}, \bar{L}_k)$. This is discussed in Section \ref{sec:trad}. However, questions about causal joint effects of treatments over time are more difficult to answer, due to the presence of time-dependent confounding. An example of a question about causal joint treatment effects is whether there is a difference in the probability of survival up to $\tau$ years had an individual been assigned by an intervention to have $A=1$ at all time points versus had they been assigned to have $A=0$ at all time points. Time-dependent confounding occurs when there are time-dependent covariates that predict subsequent treatment use, are affected by earlier treatment, and affect the outcome through pathways that are not just through subsequent treatment. The $L_k$ are time-dependent confounders in the DAG in Figure \ref{fig:dag}. The DAG could be extended in various ways, in particular so that there are long term effects of $L$ on $A$ and vice versa. For example, we could add arrows from $L_{k}$ to $A_{k+1}$ and from $A_k$ to $L_{k+2}$. Long term effects of $A$ and $L$ on survival could also be added, for example by adding arrows from $L_{k}$ and $A_{k}$ to $Y_{k+1}$.
	
	\section{Estimating treatment effects using longitudinal data}
	\label{sec:msms}
	
	\subsection{Traditional survival analysis}
	\label{sec:trad}
	
	We begin by briefly reviewing traditional methods of analysis for investigating the association between a time-dependent treatment variable and a time-to-event outcome. By far the most popular approach is Cox regression \citep{Cox:1972}. Consider a Cox regression model in which the hazard at time $t$, incorporating time-dependent covariates, is
	\begin{equation}
	\lambda(t|\bar{A}_{\lfloor t \rfloor},\bar{L}_{\lfloor t \rfloor})=\lambda_{0}(t)\exp\lb \beta_{A0} A_{\lfloor t \rfloor}+\sum_{j=1}^{\lfloor t \rfloor}\beta_{Aj} A_{\lfloor t \rfloor -j}+\sum_{j=0}^{\lfloor t \rfloor}\beta_{Lj} L_{\lfloor t \rfloor -j}\rb
	\label{eq:cox}
	\end{equation}
	where $A_{\lfloor t \rfloor}$ and $L_{\lfloor t \rfloor}$ denote the values at the most recent visit prior to time $t$, $\lambda_{0}(t)$ is the baseline hazard, and the $\beta$ parameters are log hazard ratios. The hazard ratio $\exp\lb\beta_{A0}\rb$ is the instantaneous multiplicative effect of the current treatment $A_{\lfloor t \rfloor}$ on the hazard among individuals at risk at time $t$, assumed to be the same for all $t$, adjusted for past variables (including past treatment), which are confounders of the association between $A_{\lfloor t \rfloor}$ and the current hazard. The other model parameters do not have a straightforward interpretation. For example, the coefficient for $A_{\lfloor t \rfloor-1}$, $\beta_{A1}$, is conditional on covariates that include $A_{\lfloor t \rfloor}$ and $L_{\lfloor t \rfloor}$, which are on the mediating pathway from $A_{\lfloor t \rfloor-1}$ to survival, and so its interpretation is complicated. Hence, the estimation of joint effects of treatments over time is not accommodated using the traditional Cox modelling approach with time-dependent covariates. Furthermore, a growing body of work has explained that hazard ratios do not have a straightforward causal interpretation \citep{Hernan:2010,Aalen:2015,Martinussen:2019} and so there are subtleties in the interpretation of $\beta_{A0}$ even when all confounders have been included. 
	
	Aalen's additive hazard model \citep{Aalen:1989,Aalen:2008} has been much less used in practice, but its attractive properties are increasingly being recognised \citep{Martinussen:2013}. Consider an additive hazard model in which the hazard at time $t$, incorporating time-dependent covariates, is
	\begin{equation}
	\lambda(t|\bar{A}_{\lfloor t \rfloor},\bar{L}_{\lfloor t \rfloor})=\alpha_{0}(t)+\alpha_{A0}(t) A_{\lfloor t \rfloor}+\sum_{j=1}^{\lfloor t \rfloor-1}\alpha_{Aj}(t) A_{\lfloor t \rfloor -j}+\sum_{j=0}^{\lfloor t \rfloor}\alpha_{Lj}(t) L_{\lfloor t \rfloor -j}
	\label{eq:aalen}
	\end{equation}
	where the parameters $\alpha_0(t),\alpha_{Aj}(t),\alpha_{Lj}(t)$ ($j=0,\ldots,4$) are arbitrary functions of time, meaning that the model is fully non-parametric. The results from the additive hazard model are typically presented as cumulative coefficients, e.g. $\int_{0}^{t}\alpha_{A0}(s)ds$. The discussion above about the interpretation of $\beta_{A0}$ is equally relevant to $\alpha_{A0}(t)$, and again the presence of time-dependent confounding means that joint effects of treatments over time cannot be estimated directly from the traditional additive hazard model. An advantage of the additive hazard model relative to the Cox model is that the parameters of the additive hazard model are collapsible, meaning that the parameter associated with a given covariate in a given model has the same interpretation as that in a model which is additionally adjusted for variables that are not associated with that covariate \citep{Martinussen:2013}. By contrast, hazard ratios are non-collapsible, meaning that the Cox model does not have this property. Collapsibility has implications for the relation between conditional models and marginal models. In Section 4, we use this property to show that a conditional additive hazard model, of a form such as that in (\ref{eq:aalen}), has a useful role in the simulation of longitudinal data in such a way that the form of the correctly specified MSM for the hazard is known.
	
	\subsection{Marginal structural hazard models}
	
	MSMs are models for counterfactual outcomes. We let $T^{\underline{a}_{0}}$ denote the counterfactual event time for a given individual had they followed treatment regime $\underline{a}_{0}$ from visit $0$ onwards. The marginal hazard at time $t$ under the possibly counter-to-fact treatment regime $\underline{a}_{0}$ is the hazard in the population if everyone were to receive that treatment regime, and is denoted $\lambda_{T^{\underline{a}_{0}}}(t)$. 
	
	In the context of time-to-event outcomes, the MSM is usually assumed to take the Cox proportional hazards form
	\begin{equation}
	\lambda_{T^{\underline{a}_{0}}}(t)=\lambda_{0}(t)\exp\lcb g(\bar{a}_{\lfloor t \rfloor};\tilde\beta_A)\rcb
	\label{eq:msm.cox}
	\end{equation}
	where $\lambda_{0}(t)$ is the baseline counterfactual hazard, $\bar{a}_{\lfloor t \rfloor}$ denotes treatment pattern up to the most recent visit prior to time $t$, $g(\bar{a}_{\lfloor t \rfloor};\tilde\beta_A)$ is a function (to be specified) of treatment pattern $\bar{a}_{\lfloor t \rfloor}$, and $\tilde\beta_A$ is a vector of log hazard ratios. The hazard model could take any form, however, and we also consider MSMs based on Aalen's additive hazard model:
	\begin{equation}
	\lambda_{T^{\underline{a}_{0}}}(t)=\tilde\alpha_{0}(t)+ g(\bar{a}_{\lfloor t \rfloor};\tilde\alpha_A(t))
	\label{eq:msm.aalen}
	\end{equation}
	
	The MSM must specify how the hazard at time $t$ depends on the history of treatment up to time $t$, $\bar a_{\lfloor t \rfloor}$, through the function $g(\cdot)$. In a simple form for the MSM, the hazard at time $t$ is specified to depend only on the current level of treatment, so that $g(\bar{a}_{\lfloor t \rfloor};\tilde\beta_A)=\tilde\beta_A a_{\lfloor t \rfloor}$ in the Cox MSM and $g(\bar{a}_{\lfloor t \rfloor};\tilde\alpha_A(t))=\tilde\alpha_A(t) a_{\lfloor t \rfloor}$ in the Aalen MSM. Other examples are for the hazard at $t$ to depend on duration of treatment, using $g(\bar{a}_{\lfloor t \rfloor};\tilde\beta_A)=\tilde\beta_A \sum_{j=0}^{\lfloor t \rfloor}a_{\lfloor t \rfloor -j}$ in the Cox MSM and $g(\bar{a}_{\lfloor t \rfloor};\tilde\alpha_A(t))=\tilde\alpha_A(t) \sum_{j=0}^{\lfloor t \rfloor}a_{\lfloor t \rfloor -j}$ in the Aalen MSM, or on the history of treatment through main effect terms for treatment at each visit, using $g(\bar{a}_{\lfloor t \rfloor};\tilde\beta_A)=\sum_{j=0}^{\lfloor t \rfloor}\tilde\beta_{Aj} a_{\lfloor t \rfloor-j}$ in the Cox MSM and $g(\bar{a}_{\lfloor t \rfloor};\tilde\alpha_A(t))=\sum_{j=0}^{\lfloor t \rfloor}\tilde\alpha_{Aj}(t) a_{\lfloor t \rfloor -j}$ in the Aalen MSM. 
	
	When there is confounding an MSM cannot be estimated by fitting the model to the observed data using standard regression. The most commonly used estimation approach uses IPTW, in which individuals are reweighted using time-dependent weights \citep{Daniel:2013,Cole:2008}. Further details on the weights are given in the Supplementary Materials (Section A1). MSMs can also be estimated using the g-formula \citep{Robins:1986,Daniel:2011}, and the methods described in Section \ref{sec:sim.msm} make use of this. The use of MSMs estimated using IPTW to estimate causal effects of joint treatments over time involves the four key assumptions of no interference, positivity, consistency, and conditional exchangeability (no unmeasured confounding) \citep{Robins:2000,Vanderweele:2009,Daniel:2013}. The no interference assumption is that the counterfactual event time for a given individual, $T^{\underline{a}_{0}}$, does not depend on the treatment received by any other individuals. The positivity assumption is that each individual has a strictly non-zero probability of receiving each given pattern of treatments over time. Consistency means that an individual's observed outcome is equal to the counterfactual outcome when the assigned treatment pattern is set to that which was actually received, $T_i=T_i^{\underline{A}_{0,i}}$. The conditional exchangeability assumption can be expressed formally as $T^{\bar{A}_{k-1},\underline{a}_{k}}\indep A_k|\bar{A}_{k-1},\bar{L}_{k},T\geq k$ for all feasible $\underline{a}_{k}$, where $T^{\bar{A}_{k-1},\underline{a}_{k}}$ denotes the counterfactual event time had an individual followed their observed treatment pattern up to time $k-1$, $\bar{A}_{k-1}$, and had their treatments been set to $\underline{a}_{k}$ from time $k$ onwards, given survival to time $k$. The conditional exchangeability assumption means that among individuals who remain at risk of the event at time $k$, the treatment $A_k$ received at time $k$ may depend on past treatment and covariates $\bar{A}_{k-1}$ and $\bar{L}_{k}$, but that, conditional on these, it does not depend on the remaining lifetime that would apply if all future treatments were set to any particular values $\underline{a}_{k}$.
	
	The Cox MSM gives rise to estimates of the log hazard ratios $\tilde\beta_A$, and the Aalen MSM to estimates of cumulative coefficients $\int_{0}^{t}\tilde\alpha_A(s)ds$. As noted in Section 3.1, hazard-based estimands, such as hazard ratios or differences in cumulative hazards, have been shown not to have a direct causal interpretation. Therefore, it is desirable to translate the estimates from the MSM into an estimate for a causal estimand such as a risk difference or a risk ratio, both of which are derived from survival probabilities. Based on the Cox MSM in (\ref{eq:msm.cox}), the counterfactual survival probability at time $t$ is. 
	\begin{equation}
	\begin{split}
	\Pr(T^{\underline{a}_{0}}>t)&=\exp\lb-e^{g(a_{0};\tilde\beta_A)}\int_{0}^{1}\lambda_{0}(s)ds-e^{ g(\bar{a}_{1};\tilde\beta_A)}\int_{1}^{2}\lambda_{0}(s)ds\cdots -e^{g(\bar{a}_{\lfloor t \rfloor};\tilde\beta_A)}\int_{\lfloor t \rfloor}^{t}\lambda_{0}(s)ds\rb
	\end{split}
	\end{equation}
	where the baseline cumulative hazard can be estimated using (an inverse probability weighted) Breslow's estimator. The counterfactual survival probability based on the Aalen MSM in (\ref{eq:msm.aalen}) is
	
	\begin{equation}
	{\small
		\Pr(T^{\underline{a}_{0}}>t)=\exp\left(-\int_{0}^{t}\tilde\alpha_{0}(s)ds-\int_{0}^{1}g(a_{0};\tilde\alpha_A(s))ds-\int_{1}^{2}g(\bar{a}_{1};\tilde\alpha_A(s))ds\cdots -\int_{\lfloor t \rfloor}^{t}g(\bar{a}_{\lfloor t \rfloor};\tilde\alpha_A(s))ds\right)
		\label{eq:surv.aalen.msm}}
	\end{equation}
	
	\section{Simulation from MSMs}
	\label{sec:sim.msm}
	
	As noted in Section 1, when conducting a simulation study to evaluate and compare the properties of analysis methods, it is important to be able to generate the data in such a way that the forms of any models to be estimated using the simulated data are known based on the data generating mechanism. In our context, for evaluations involving MSMs it is therefore important to know the correct form of the MSM, and hence know or be able to derive the true values of its parameters and of causal estimands of interest such as risk differences or risk ratios. It may also be of interest in some contexts to evaluate the impact of using a mis-specfied MSM, in which case we need to understand how the model under consideration differs from the correctly specified MSM. 
	
	When simulating longitudinal and time-to-event data, as depicted in the DAG in Figure 1, it is natural to generate the data sequentially in time. We provide a detailed algorithm in Section 5. Briefly, the procedure starts by generating $U$, then $L_0|U$, then $A_0|L_0,U$, and then event times in period $0<t<1$ using the hazard $\lambda(t|A_0,L_0,U)$. The next step is to generate $L_1|L_0,A_0,U,T\geq1$, followed by $A_1|L_0,L_1,A_0,U,T\geq1$, and then event times in period $1\leq t<2$ using the hazard $\lambda(t|A_0,A_1,L_0,L_1,U)$. Analogous steps are then carried out for each of visits 2, 3 and so on up to $K$. This procedure uses the conditional hazards $\lambda(t|\bar A_{\lfloor t \rfloor}=\bar a_{\lfloor t \rfloor},\bar L_{\lfloor t \rfloor},U)$. The MSM describes instead the marginal hazard, which is a function only of the assigned treatment up to time $t$, $\bar{a}_{\lfloor t \rfloor}$, and not of $\bar{L}_{\lfloor t \rfloor}$ or on $U$. The question therefore arises as to what the form of the MSM is under the sequential data generating procedure outlined above, which uses a conditional hazard model and conditional models for the time-dependent covariates. 
	
	\subsection{Link between conditional and marginal hazard models}
	
	In this section we derive general results for the link between the conditional models used to simulate the longitudinal and time-to-event data and the MSM $\lambda_{\Tc}(t)$. These general results are then used in the context of additive hazard models and Cox models. This extends some of the work of \cite{Martinussen:2013} to the longitudinal setting. Our overall approach is to first use the g-formula for time-to-event outcomes \citep{Robins:1986,Keil:2014,Daniel:2013} to express the survivor function for counterfactual event times, $\Pr(T^{\underline{a}_0}\geq t)$, in terms of conditional distributions of observed event times and variables $A,L,U$, and then use the fact that the hazard can be expressed as minus the derivative of the log of the survivor function: $\lambda_{T^{\underline{a}_0}}(t)=\frac{-\frac{d}{dt}\Pr(T^{\underline{a}_0}\geq t)}{\Pr(T^{\underline{a}_0}\geq t)}$. We first consider the effect of treatment at time 0, $a_0$, on the hazard at times $0<t<1$, and then the effect of treatment at times 0 and 1 on the hazard at times $1\leq t<2$, and so on. 
	
	By averaging over $L_0$ and $U$, the marginal survival probability $\Pr(T^{\underline{a}_0}\geq t)$ for $0<t<1$ can be expressed as
	\begin{equation}
	\begin{split}
	\Pr(T^{\underline{a}_0}\geq t)=&\int\Pr(T^{\underline{a}_0}\geq t|L_0,U)f(L_0,U)dL_0dU\\
	=&\int\Pr(T\geq t|A_0=a_0,L_0,U)f(L_0,U)dL_0dU\\
	\end{split}
	\label{eq:margsurv.t1}
	\end{equation}
	where the second line follows from the conditional exchangeability assumption $T^{\underline{a}_0}\indep A_0|L_0$ and consistency. Using the relation between the hazard and the survivor function the hazard corresponding to the survival function in (\ref{eq:margsurv.t1}) can be written
	\begin{equation}
	\begin{split}
	\lambda_{T^{\underline{a}_0}}(t)=&\frac{-\int\frac{d}{dt}\Pr(T\geq t|A_0=a_0,L_0,U)f(L_0,U)dL_0dU}{\int\Pr(T\geq t|A_0=a_0,L_0,U)f(L_0,U)dL_0dU}\\
	=&\frac{\int\lambda(t|A_0=a_0,L_0,U)\Pr(T\geq t|A_0=a_0,L_0,U)f(L_0,U)dL_0dU}{\int\Pr(T\geq t|A_0=a_0,L_0,U)f(L_0,U)dL_0dU}\\
	=&\frac{E_{L_0,U}\left\{\lambda(t|A_0=a_0,L_0,U)\Pr(T\geq t|A_0=a_0,L_0,U)\right\}}{E_{L_0,U}\left\{\Pr(T\geq t|A_0=a_0,L_0,U)\right\}}\\
	=&\frac{E_{L_0,U}\left\{\lambda(t|A_0=a_0,L_0,U)\exp\lb-\int_{0}^{t}\lambda(s|A_0=a_0,L_0,U)ds\rb\right\}}{E_{L_0,U}\left\{\exp\lb-\int_{0}^{t}\lambda(s|A_0=a_0,L_0,U)ds\rb\right\}}\\
	\end{split}
	\label{eq:marghaz.t1}
	\end{equation}
	where $E_{L_0,U}(\cdot)$ denotes the expectation over the joint distribution of $L_0$ and $U$. For $0<t<1$, the MSM $\lambda_{T^{\underline a_0}}(t)$ can therefore be expressed as a function of the conditional hazard $\lambda(t|A_0=a_0,L_0,U)$ and conditional distributions of variables $A_0,L_0,U$.
	
	Next, we derive an expression for the marginal survivor function $\Pr(T^{\underline{a}_0}\geq t)$ for $1\leq t<2$, followed by an expression for the corresponding hazard. To derive the survivor function, first consider averaging over the baseline variables $L_0$ and $U$. This gives
	\begin{equation}
	\begin{split}
	\Pr(T^{\underline{a}_0}\geq t)=&\int\Pr(T^{\underline{a}_0}\geq t|L_0,U)f(L_0,U)dL_0dU\\
	=&\int\Pr(T^{\underline{a}_0}\geq t|A_0=a_0,L_0,U)f(L_0,U)dL_0dU\\
	\end{split}
	\end{equation}
	where the second line follows from the conditional exchangeability assumption $T^{\underline{a}_0}\indep A_0|L_0$ and consistency. Because here our focus is on $1\leq t<2$, the above can be written 
	\begin{equation}
	\begin{split}
	\Pr(T^{\underline{a}_0}\geq t)=&\int\Pr(T^{\underline{a}_0}\geq t|A_0=a_0,L_0,U,T^{\underline{a}_0}\geq 1)\Pr(T^{\underline{a}_0}\geq 1|A_0=a_0,L_0,U)f(L_0,U)dL_0dU\\
	=&\int\Pr(T^{\underline{a}_0}\geq t|A_0=a_0,L_0,U,T\geq 1)\Pr(T\geq 1|A_0=a_0,L_0,U)f(L_0,U)dL_0dU
	\end{split}
	\end{equation}
	where the second line follows because the events that $T^{\underline{a}_0}\geq 1$ and  $T\geq 1$ are the same for individuals with $A_0=a_0$. In the next step we first average over $L_1|L_0,U,T\geq 1$ and then use the conditional exchangeability assumption $T^{\underline{a}_0}\indep A_1|\bar L_1,A_0=a_0,T\geq 1$ and consistency to give
	{\footnotesize
		\begin{equation}
		\begin{split}
		\Pr(T^{\underline{a}_0}\geq t)=&\int\Pr(T^{\underline{a}_0}\geq t|A_0=a_0,\bar L_1,U,T\geq 1)\Pr(T\geq 1|A_0=a_0,L_0,U)\\
		&\times f(L_1|A_0=a_0,L_0,U,T\geq 1)f(L_0,U)d\bar L_1dU\\
		=&\int\Pr(T\geq t|A_0=a_0,A_1=a_1,\bar L_1,U,T\geq 1)\Pr(T\geq 1|A_0=a_0,L_0,U)\\
		&\times f(L_1|A_0=a_0,L_0,U,T\geq 1)f(L_0,U)d\bar L_1dU\\
		=&E_{L_0,U}\lsb E_{L_1|A_0=a_0,L_0,U,T\geq 1}\left\{\Pr(T\geq t|\bar A_1=\bar a_1,\bar L_1,U,T\geq 1)\Pr(T\geq 1|A_0=a_0,L_0,U)\right\}\rsb\\
		\end{split}
		\end{equation}}
	Finally, using the relation between the hazard and survivor function it can be shown that for $1\leq t<2$
	{\footnotesize
		\begin{equation}
		\lambda_{T^{\underline{a}_0}}(t)=\frac{E_{L_0,U}\lsb E_{L_1|A_0=a_0,L_0,U,T\geq 1}\lcb\lambda(t|\bar A_1=\bar a_1,\bar L_1,U)\exp\lb-\int_{0}^{1}\lambda(s|A_0=a_0,L_0,U)ds-\int_{1}^{t}\lambda(s|\bar A_1=\bar a_1,\bar L_1,U)ds\rb \rcb\rsb}{E_{L_0,U}\lsb E_{L_1|A_0=a_0,L_0,U,T\geq 1}\lcb\exp\lb-\int_{0}^{1}\lambda(s|A_0=a_0,L_0,U)ds-\int_{1}^{t}\lambda(s|\bar A_1=\bar a_1,\bar L_1,U)ds\rb\rcb\rsb}\\
		\label{eq:marghaz.t2}
		\end{equation}}
	It follows that for $1\leq t<2$ the MSM $\lambda_{T^{\underline a_0}}(t)$ can be expressed as a function of the conditional hazard $\lambda(t|\bar{A}_{1},\bar{L}_{1},U)$ and conditional distributions of variables $\bar A_1,\bar L_1,U$.
	
	A general expression for the MSM at times $k\leq t<k+1$ is 
	\begin{equation}
	\lambda_{T^{\underline{a}_0}}(t)=\frac{E_{L_0,U}\lsb E_{L_1|A_0=a_0,L_0,U,T\geq 1}\lcb\cdots E_{L_{k}|\bar{A}_{k-1}=\bar{a}_{k-1},\bar{L}_{k-1},U,T\geq k}\lb\lambda(t|\bar A_k=\bar a_k,\bar L_k,U,)\mathcal{Q}_k\rb \rcb\rsb}{E_{L_0,U}\lsb E_{L_1|A_0=a_0,L_0,U,T\geq 1}\lcb\cdots E_{L_{k}|\bar{A}_{k-1}=\bar{a}_{k-1},\bar{L}_{k-1},U,T\geq k}\lb\mathcal{Q}_k\rb \rcb\rsb}\\
	\end{equation}
	where $\mathcal{Q}_k=\prod_{j=0}^{k-1}\exp\lb-\int_{j}^{j+1}\lambda(s|\bar{A}_{j}=a_{j},\bar{L}_{j},U)ds\rb \exp\lb-\int_{k}^{t}\lambda(s|\bar{A}_{k}=a_{k},\bar{L}_{k},U)ds\rb$. 
	
	The above results show how the MSM $\lambda_{\Tc}(t)$ can be expressed in terms of the conditional hazard model for the observed data, $\lambda(t|\bar{A}_{\lfloor t \rfloor},\bar{L}_{\lfloor t \rfloor},U)$, and conditional distributions for the observed time-dependent covariates. The results were derived by making use of the g-formula. We next apply these results to the situations in which the conditional hazard model $\lambda(t|\bar A_{\lfloor t \rfloor},\bar L_{\lfloor t \rfloor},U)$ follows an Aalen additive hazard model or a Cox model.
	
	\subsection{Results using conditional additive hazard models}
	
	Suppose that the conditional hazard model is of the additive form
	\begin{equation}
	\lambda(t|\bar A_{\lfloor t \rfloor},\bar L_{\lfloor t \rfloor},U)=\alpha_0(t)+\alpha_A^{\top}(t)v(\bar A_{\lfloor t \rfloor})+\alpha_L^{\top}(t)w(\bar L_{\lfloor t \rfloor})+\alpha_U(t)U
	\label{eq:addhaz.cond}
	\end{equation}
	where $\alpha_A(t)$ and $\alpha_L(t)$ are vectors of parameters and the hazard at time $t$ depends on a known vector function of $\bar A_{\lfloor t \rfloor}$, $v(\bar A_{\lfloor t \rfloor})$, and a known vector function of $\bar L_{\lfloor t \rfloor}$, $w(\bar L_{\lfloor t \rfloor})$. 
	
	It can be shown that $\lambda_{\Tc}(t)$ also takes the form of an additive hazard model in this case. We provide results for $0<t<1$ and $1\leq t<2$ to illustrate the point. For $0< t<1$, using the general expression in (\ref{eq:marghaz.t1}), we have
	\begin{equation}
	{\small
		\lambda_{\Tc}(t)= \alpha_0(t)+\alpha_A^{\top}(t)v(a_0)+\frac{E_{L_0,U}\left\{ \left(\alpha_L^{\top}(t)w(L_0)+\alpha_U(t)U\right)\exp\lb-\int_{0}^{t}(\alpha_L^{\top}(s)w(L_0)+\alpha_U(s)U)ds\rb\right\}}{E_{L_0,U}\left\{ \exp\lb-\int_{0}^{t}(\alpha_L^{\top}(s)w(L_0)+\alpha_U(s)U)ds\rb\right\}}}
	\label{eq:marhaz.aalen.t1}
	\end{equation}
	This expression for $\lambda_{\Tc}(t)$ ($0<t<1$) is of the additive form, $\lambda_{\Tc}(t)=\tilde{\alpha}_0(t)+\alpha^{\top}_A(t)v(a_0)$. The coefficient for $v(a_0)$, $\alpha_A(t)$, is the same as in the conditional hazard model, whereas the intercept $\tilde{\alpha}_0(t)$ is now the sum of $\alpha_0(t)$ and the third term in the expression in (\ref{eq:marhaz.aalen.t1}). Note that since the treatment is binary $v(a_0)=a_0$. The result in (\ref{eq:marhaz.aalen.t1}) is similar to that derived by \cite{Martinussen:2013}, who considered the form of the marginal hazard in the setting of a point treatment, except they did not incorporate a $U$ variable. 
	
	For $1\leq t<2$ it can be shown using (\ref{eq:marghaz.t2}) that the MSM is of the form
	\begin{equation}
	{\small
		\begin{split}
		\lambda&_{\Tc}(t)= \alpha_0(t)+\alpha_A^{\top}(t)v(\bar{a}_1)+\frac{E_{L_0,U}\lsb E_{L_1|A_0=a_0,L_0,U,T\geq 1}\lcb (\alpha_L^{\top}(t)w(\bar L_1)+\alpha_U(t)U) \mathcal R(\bar L_1, U)\rcb\rsb}{E_{L_0,U}\lsb E_{L_1|A_0=a_0,L_0,U,T\geq 1}\lcb  \mathcal R(\bar L_1, U)\rcb\rsb}
		\end{split}}
	\label{eq:marhaz.aalen.t2}
	\end{equation}
	where  $\mathcal R(\bar L_1, U)=\exp\lb-\int_{0}^{1}(\alpha_L^{\top}(s)w(L_0)+\alpha_U(s)U)ds-\int_{1}^{t}(\alpha_L^{\top}(s)w(\bar{L}_1)+\alpha_U(s)U)ds\rb$. The third term of (\ref{eq:marhaz.aalen.t2})  is a function of $a_0$. It follows from this expression that for a binary treatment the MSM $\lambda_{\Tc}(t)$ is of the additive hazard form $\lambda_{\Tc}(t)=\tilde{\alpha}_0(t)+\alpha^{\top}_A(t)v(\bar a_1)+\tilde{\alpha}^*_A(t)a_0$. In the setting where $\alpha_A^{\top}(t)v(\bar{a}_1)=\alpha_{A0}(t)a_1+\alpha_{A1}(t)a_0$, the coefficient for $a_1$ in the MSM $\lambda_{\Tc}(t)$ ($1\leq t<2$) is the same as that in the conditional hazard model, $\alpha_{A0}(t)$, whereas the intercept and the coefficient for $a_0$ are different from those in the conditional hazard model. 
	
	Closed-form expressions for the third terms in (\ref{eq:marhaz.aalen.t1}) and (\ref{eq:marhaz.aalen.t2}) (ratios of nested expectations) can be derived for special cases. In particular, if $L_0,U$ have a bivariate normal distribution, and the distribution of $L_1|A_0=a_0,L_0,U,T\geq 1$ is normal, then the expections can be evaluated using the properties of the Laplace transform, or equivalently the moment generating function for the normal distribution. We provide expressions for this special case in the Supplementary Materials (Section A2). However, the result that the MSM  $\lambda_{\Tc}(t)$ is of an additive form when the conditional hazard model is an additive model does not rely on distributional assumptions for $L$ and $U$. In Section 6.2 we describe an alternative general approach to deriving the true values of the parameters of the MSM through simulation.
	
	In the conditional additive hazard model in (\ref{eq:addhaz.cond}) the treatment history is included in the general form $v(\bar A_t)$. In practice, as discussed in Section 3.2, this form has to be specified. Suppose that the conditional hazard model was of a form such that the hazard at time $t$ depends only on the current treatment status $A_\lfloor t\rfloor$, i.e. $\lambda(t|\bar A_{\lfloor t \rfloor},\bar L_{\lfloor t \rfloor},U)=\alpha_0(t)+\alpha_A(t)A_{\lfloor t \rfloor}+\alpha_L^{\top}(t)w(\bar L_{\lfloor t \rfloor})+\alpha_U(t)U$. The result in (\ref{eq:marhaz.aalen.t2}) shows that even if the conditional hazard at time $t$ ($1\leq t<2$) depends on treatment only through the current level, $a_1$, the MSM depends on both $a_0$ and $a_1$ for $1\leq t<2$. The intuition behind this result is that $A_0$ affects $L_1$ and hence after the averaging over $L_1$, the marginal hazard at time $t$ ($1\leq t<2$) depends on $a_0$. In general, even if the conditional hazard at time $t$ depends on treatment only through the current level, $a_{\lfloor t \rfloor}$, the MSM depends on the whole history of treatment $\bar a_{\lfloor t \rfloor}$. In the Supplementary Material (Section A3) we extend the results to the setting where the conditional hazard model (\ref{eq:addhaz.cond}) additionally includes interactions between $\bar A_{\lfloor t \rfloor}$ and $\bar L_{\lfloor t \rfloor}$. 
	
	\subsection{Results using conditional Cox models}
	
	Suppose instead that the conditional hazard model is of the Cox proportional hazards form
	\begin{equation}
	\lambda(t|\bar A_{\lfloor t \rfloor},\bar L_{\lfloor t \rfloor},U)=\lambda_0(t) \exp\lb\beta_A^{\top}v(\bar A_{\lfloor t \rfloor})+\beta_L^{\top}w(\bar L_{\lfloor t \rfloor})+\beta_U U\rb
	\end{equation}
	For $0< t<1$, using the general expression in (\ref{eq:marghaz.t1}), the MSM takes the form
	\begin{equation}
	{\small
		\lambda_{\Tc}(t)= \lambda_0(t) \exp\lb\beta_A^{\top}v(a_0)\rb\lsb\frac{E_{L_0,U}\lcb \exp\lb\beta_L^{\top}w(L_0)+\beta_U U\rb\exp\lb-\int_{0}^{t}\lambda_0(s) e^{\beta_A^{\top}v(a_0)+\beta_L^{\top}w(L_0)+\beta_U U} ds\rb\rcb}{E_{L_0,U}\lcb \exp\lb-\int_{0}^{t}\lambda_0(s) e^{\beta_A^{\top}v(a_0)+\beta_L^{\top}w(L_0)+\beta_U U} ds\rb\rcb}\rsb
		\label{eq:marhaz.cox.t1}}
	\end{equation}
	The ratio of expectations in the third term in the above expression is a complicated function of both $t$ and $a_0$, and $\lambda_{\Tc}(t)$ no longer takes the Cox model form. A closed form expression for the third term of (\ref{eq:marhaz.cox.t1}) is not generally available, even in the setting of bivariate normality for $L_0,U$. 
	
	Similar results to those provided here for the Cox model were derived by \cite{Young:2014}, who focused on a discrete time setting. We discuss their results further in Section \ref{sec:discussion}.
	
	\subsection{Summary}
	
	When the conditional hazard model $\lambda(t|\bar A_{\lfloor t \rfloor},\bar L_{\lfloor t \rfloor},U)$ is additive, we have shown that the MSM $\lambda_{\Tc}(t)$ is also additive. The coefficients for $\bar a_t$ in the MSM differ from those in the conditional model except for $0<t<1$ - that is, except up to visit $k=1$. The intercepts in the conditional model differ from those in the MSM at all time points. Even if the conditional hazard model depends on treatment only through the current level, the MSM depends on the whole treatment history. 
	
	When the conditional hazard model is a Cox model, the MSM is no longer a Cox model; instead it takes a complex form with the effect of $\bar a_k$ on the hazard being a complex function of time.
	
	\section{Simulation algorithm}
	\label{sec:alg}
	
	It follows from the results of Section \ref{sec:sim.msm} that if longitudinal data are simulated according to a conditional additive hazard model, then the marginal hazard model used in a MSM analysis is also additive and hence can be correctly specified. In this section we describe an example simulation algorithm which results in a known additive form for the MSM. This is intended as a particular illustration of a general approach and the algorithm can easily be modified for other data-generating mechanisms. In Section 6 we illustrate the practical implementation of the algorithm, and R code is provided at https://github.com/ruthkeogh/causal\_sim. 
	
	Longitudinal data are generated at 5 visits $k=0,\ldots,4$ for a single time-dependent continuous variable $L$, for example representing a biomarker, and for a binary treatment $A$ and continuous variable $U$, representing an individual frailty term. The example algorithm uses a conditional hazard of the form $\lambda(t|\bar A_{\lfloor t \rfloor},\bar L_{\lfloor t \rfloor},U)=\alpha_0+\alpha_A A_{\lfloor t \rfloor}+\alpha_L L_{\lfloor t \rfloor}+\alpha_U U$. Here we focus on constant conditional baseline hazard and constant coefficients, which simplifies the generation of event times. An extension of the algorithm to accommodate more complex forms for the hazard is described in the Supplementary Materials (Section A4), and is based on generating event times from a piecewise exponential distribution. The conditional hazard at time $t$ depends on the current values of $A$ and $L$, but not on past values. The implied form of the MSM is $\lambda_{\Tc}(t)=\tilde{\alpha}_{0}(t)+\sum_{j=0}^{\lfloor t \rfloor}\tilde{\alpha}_{Aj}(t)a_{\lfloor t \rfloor-j}$. In the example algorithm, higher values of the biomarker $L$ are associated with higher propensity to receive the treatment and higher hazard. The biomarker also increases with time. The treatment lowers the value of $L$ and lowers the hazard. Event times are generated in the range $0<T\leq 5$ and there is administrative censoring at time 5. 
	
	The steps to generate the longitudinal data are as follows for each individual $i=1,\ldots,n$:
	
	\begin{enumerate}
		\item Generate the individual frailty term $U$ from a normal distribution with mean 0 and standard deviation $0.1$.
		\item Generate $L_0$ from a normal distribution with mean $U$ and standard deviation $1$.
		\item Generate $A_0$ from a Bernoulli distribution with $\mathrm{logit } \Pr(A_0=1|L_0)=-2+0.5L_0$.
		\item The conditional hazard is $\lambda(t|\bar A_{\lfloor t \rfloor},\bar L_{\lfloor t \rfloor},U)=0.7-0.2 A_{\lfloor t \rfloor}+0.05L_{\lfloor t \rfloor}+0.05U$. Event times are generated in the period $0<t<1$ as follows. First generate $V\sim \mathrm{Uniform}(0,1)$ and calculate $T^*=-\log(V)/\lambda(t|A_0,L_0,U)$. If $T^*<1$ the event time is set to be $T=T^*$. Individuals with $T^*>1$ remain at risk of the event at time $t=1$.\\
		\\
		For individuals who remain at risk of the event at visit time $k=1$:
		\item Generate $L_k$ from a normal distribution with mean $0.8L_{k-1}-A_{k-1}+0.1k+U$ and standard deviation $1$. 
		\item Generate $A_k$ from a Bernoulli distribution with $\mathrm{logit } \Pr(A_k=1|\bar A_{k-1},\bar L_k,T\geq k)=-2+0.5L_k+A_{k-1}$.
		\item Generate event times in the period $k\leq t< k+1$. First generate $V\sim \mathrm{Uniform}(0,1)$ and calculate $T^*=-\log(V)/\lambda(t|\bar{A}_k,\bar{L}_k,U)$. If $T^*<1$ the event time is set to be $T=k+T^*$. Individuals with $T^*>1$ remain at risk of the event at time $k+1$.
		\item Repeat steps 5-7 for $k=2,3,4$. Individuals who do not have an event time generated in the period $0<t<5$ are administratively censored at time 5. 
	\end{enumerate}
	
	\section{Simulation illustration}
	\label{sec:illustration}
	
	\subsection{Methods and estimands}
	
	We illustrate the algorithm described in Section 5 by generating $1000$ simulated data sets for each of $n=5000$ individuals. An MSM is fitted to each simulated data set using IPTW (MSM-IPTW). The correctly specified MSM is of the form $\lambda_{\Tc}(t)=\tilde{\alpha}_{0}(t)+\sum_{j=0}^{\lfloor t \rfloor}\tilde{\alpha}_{Aj}(t)a_{\lfloor t\rfloor -j}$. Stabilized weights were used for the IPTW estimation and the weights were estimated using logistic regression, with $\mathrm{logit }\Pr(A_k=1|\bar{A}_{k-1},T\geq k)=\gamma_0+\gamma_A A_{k-1}$ and $\mathrm{logit }\Pr(A_k=1|\bar{L}_{k},\bar{A}_{k-1},T\geq k)=\gamma_0+\gamma_A A_{k-1}+\gamma_L L_{k}$ (see Supplementary Materials Section A1). The propensity score models are correctly specified according the data generation mechanism.
	
	The estimands of interest are the cumulative coefficients $\int_{0}^{t}\tilde{\alpha}_{0}(s)ds$ and $\int_{0}^{t}\tilde{\alpha}_{Aj}(s)ds$ ($j=0,1,2,3,4$) and marginal survival probabilities for two treatment regimes: `always treated' ($\Pr(T^{\underline{a}_0=1}>t)$) or `never treated' ($\Pr(T^{\underline{a}_0=0}>t)$). For each estimand we present the mean value of the estimates across simulations at times $1,2,3,4,5$ and the corresponding bias. We also obtained the empirical standard errors of the estimates as the standard deviation of the estimates across simulations at times $1,2,3,4,5$. For the bias we obtained Monte Carlo standard errors \citep{Morris:2019}. Results are also shown graphically across all time points. We expect the estimates from the MSM to be approximately unbiased, because according to our theoretical results the MSM is correctly specified. 
	
	\subsection{Obtaining true values}
	
	To calculate the bias we need to know the true values of the estimands. Closed form expressions could be derived for the parameters of the MSM, $\tilde{\alpha}_{0}(t)$ and $\tilde{\alpha}_{Aj}$ ($j=0,\ldots,4$), using the results given in the Supplementary Materials (Section A2), because in the data generating procedure $L_k$ and $U$ are normally (or conditionally normally) distributed. However, the results in Supplementary Materials Section A2 show that even in this relatively simple setting, the expressions for the true values of the parameters beyond time $t=1$ are complicated and it would be easy to make an error. Furthermore, when $L_k$ is a vector, for non-normally distributed $L$ or $U$, and when the form of the conditional additive hazard model is more complex, finding expressions for the parameters of the additive MSM becomes intractable. 
	
	We therefore obtain the true values of the estimands of interest using an alternative approach. This is to generate longitudinal data in a similar way to that described in the algorithm but for a large `randomized controlled trial' (RCT) where the relationships between the variables are the same as in the observational study (Figure 1), with the exception that $L_k$ does not affect $A_k$. Instead, $A_k$ is set by intervention to the fixed value determined by the treatment regime. With 5 visit times and a binary treatment, there are $2^5=32$ possible longitudinal treatment regimes. We generated trial data with $m=1000$ individuals assigned to each of the 32 possible treatment regimes. The 1000 values of $L_0$ were generated once and set to be the same in each regime. Each trial therefore contains in total $32,000$ individuals. We simulated $1000$ trials. The correctly specified MSM was fitted in each simulated trial data set without any weights - since there is no time-dependent confounding in the trial data there is no need for any weights. This provides estimates of the cumulative coefficients $\int_{0}^{t}\tilde{\alpha}_{0}(s)ds$, $\int_{0}^{t}\tilde{\alpha}_{Aj}(s)ds$, $j=0,\ldots,4$. Estimates of marginal survival probabilities in the `always treated' and `never treated' groups were obtained using (\ref{eq:surv.aalen.msm}). Note that the survival probabilities could in fact have been directly estimated using simple proportions from the RCT data, since there is only administrative censoring. This is shown in the example code provided. The true values of the estimands were taken to be the average of the estimates obtained from the large randomized trials across the 1000 simulated data sets. 
	
	\subsection{Results}
	
\begin{figure}[!h]
		\caption{Cumulative coefficients: true values, estimates obtained using MSM-IPTW from 1000 simulated data sets (faded grey lines), and mean estimated cumulative coefficients using MSM-IPTW.}
		\label{fig:coef.plot.conts}
		\centering
		\includegraphics[scale=0.4]{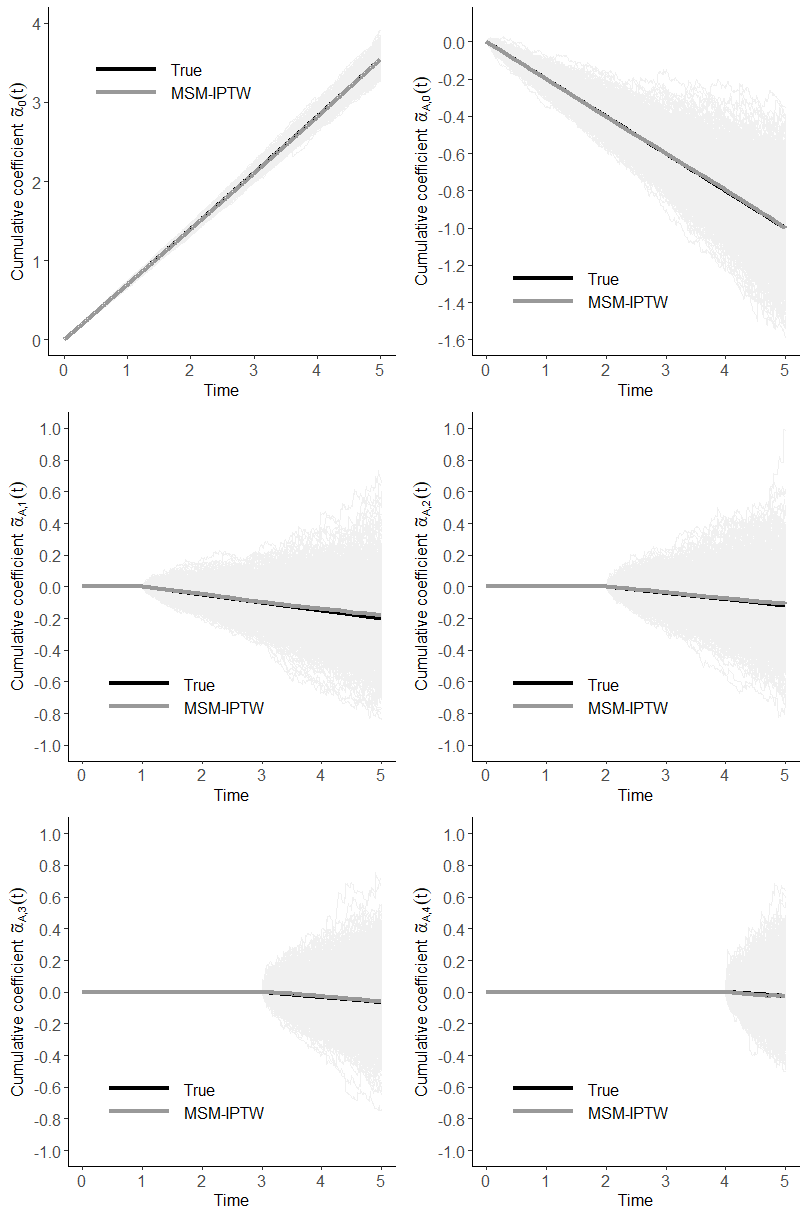}
	\end{figure}

\begin{figure}[!h]
		\caption{Survival curves for the treatment regimes `never treated' and `always treated': true survival curves, estimated survival curves obtained using MSM-IPTW from 1000 simulated data sets (faded grey lines), and the mean estimated survival curves using MSM-IPTW.}
		\label{fig:surv.plot.conts}
		\centering
		\includegraphics[scale=0.4]{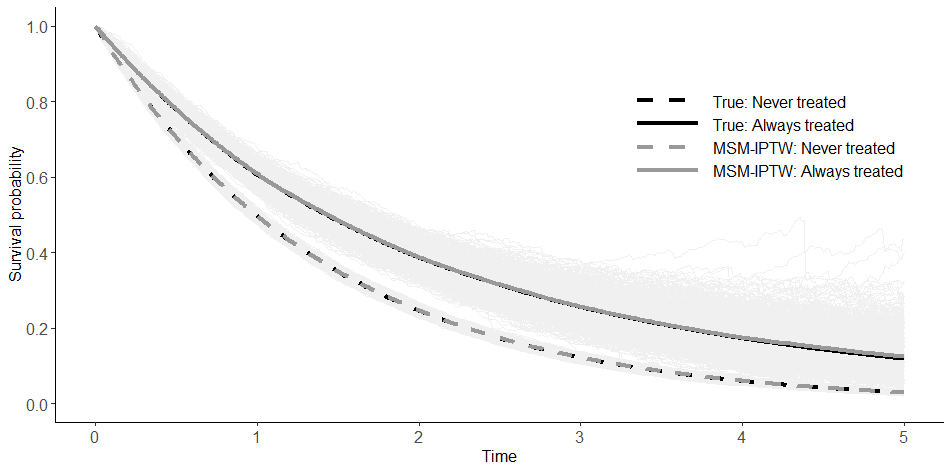}
	\end{figure}

	The results from the simulation illustration are shown in Tables \ref{tab:cumcoef} and \ref{tab:surv} and Figures \ref{fig:coef.plot.conts} and \ref{fig:surv.plot.conts}. The estimated cumulative coefficients from the MSM are approximately unbiased. The small bias in some of the cumulative coefficients is thought to be due to finite sample bias, and the plots show that it is negligible. The same applies for the survival probabilities under the `always treated' and `never treated' regimes, which are derived from the cumulative coefficients. The cumulative coefficients are imprecisely estimated, resulting in a large pointwise confidence intervals for the survival curves.
	
	\section{Discussion}
	\label{sec:discussion}
	
	In this paper we have provided results on the link between the conditional models used in the simulation of longitudinal and time-to-event data and the MSMs used in causal inference investigations to estimate the marginal effects of longitudinal treatment regimes on time-to-event outcomes. We have shown that when data are generated under an additive conditional hazard model, the form of the MSM is also additive. By contrast, when data are generated under a conditional Cox model, the form of the MSM is not a Cox model and in fact takes a complex non-standard form. We have outlined a simulation algorithm for longitudinal and time-to-event data based on the additive hazard model, and provided example simulation results to support the algebraic results. 
	
	Our results and simulation algorithm will help other researchers in the conduct of simulation studies to assess performance of methods under different conditions and to compare properties of different methods. Assessment and comparison of causal inference methods is rarely happening up to now and some comparisons are flawed. \cite{Karim:2018} compared results from an analysis using a Cox MSM with an alternative sequential Cox approach described by \cite{Gran:2010}. However they compared estimands (hazard ratios) from a marginal model with those from a conditional model, concluding incorrectly that the sequential Cox approach provides biased estimates. \cite{Gran:2019} pointed out that \cite{Karim:2018} had not made a fair comparison of the two approaches, firstly because they compared marginal with conditional estimands and secondly because the data generating procedure did not ensure that models were correctly specified under the two approaches. 
	
	The results in Section 4 were derived using the g-formula to express the MSM in terms of conditional models for the observed data. As noted in Sections 1 and 3.2, MSMs can be estimated using observed data using IPTW or the g-formula, under the assumptions outlined in Section 3.2. In the simulation illustration in Section 6 we focused on the IPTW approach, which is the most popular \citep{Clare:2018}. The general results can also be used to ascertain the form of the correctly specified MSM when using a g-formula analysis with particular specifications for the conditional models. In future work it would be of interest to compare the efficiency of estimates of survival probabilities (for example) obtained using MSMs estimated using IPTW and using the g-formula. Our simulation algorithm could be employed for this purpose, and would enable us to ensure that all models used in the analyses were correctly specified according to the data generating mechanism, including the MSM, the conditional models used in the g-formula analysis, and the propensity score models used in the IPTW analysis. 
	
	Our results also highlight the benefits of the additive hazard model for use in causal inference research, which result from its collapsibility property. More causal inference methods are emerging that make use of the additive hazard model for this reason, for example \cite{Seaman:2019,Ryalen:2019,Aalen:2019}. Our work adds to earlier results on how to simulate from MSMs in the setting of longitudinal and time-to-event data by \cite{havercroft:2012,Young:2010} and \cite{Young:2014}, who all focused on proportional hazards models. The approach of \cite{havercroft:2012} was restricted to a setting similar to that depicted in our DAG in Figure \ref{fig:dag}, but with the direct arrows from $L_k$ to $Y_{k+1}$ omitted, which is likely to be unrealistic for most purposes. Their algorithm does not generate the data depicted in the DAG in the natural sequential way. \cite{Young:2014} provided similar general results for the Cox model to those given in Section 4.3. They showed that the form of the MSM can be derived under certain conditions. Their results focused on a situation in which the conditional hazard at time $t$ depends on $A_{\lfloor t \rfloor}$, $A_{\lfloor t \rfloor -1}$ and $L_{\lfloor t \rfloor}$, but not on the further history of covariates, and in which the distribution of $L_{k+1}$ depends on $A_k$ but not on previous $L$ or the further history of $A$. Certain results also required a probit model for the conditional distribution of $L$ or the assumption that the event of interest is rare. The earlier work of \cite{Young:2010} derived data generating conditions under which a Cox MSM, a structural nested cumulative failure time model \citep{Picciotto:2012} and a structural nested accelerated failure time model \citep{Robins:1992Bka} can coincide, enabling fair comparison of the three approaches. 
	
	While the linear form of the additive hazard model brings advantages, there are also drawbacks. The additive hazard model does not restrict the hazard to be non-negative, which in turn can result in survival probabilities derived from the fitted hazard model being greater than 1. Simulation investigations for this paper showed that it is not difficult to choose a data generating procedure that gives rise to negative hazards. Researchers using this approach should therefore take care that their simulation procedure does not give negative hazards. We focused in this paper on a simplified setting with no loss-to-follow-up except through adminstrative censoring. The results extend directly to a setting with random censoring. Informative censoring can be handled through inverse probability of censoring weights, which are multiplied together with the inverse probability of treatment weights when fitting the MSM using IPTW. It is straightforward to extend our simulation algorithm to incorporate more than one $L$ variable, and even to more than one treatment variable. We focused on a binary treatment, though the results extend in theory to continuous treatments (e.g. dose). However, estimating MSMs using IPTW is not generally recommended for use with continuous exposures, since it is difficult to specify a correct distribution for the continuous treatment and even mild incorrect specification of the weights model can have significant impact on estimates \citep{Goetgeluk:2008,Naimi:2014}. Finally, we focused on a setting in which the visits times are regular and the same for all individuals. This is not representative of many of the observational data sets faced in practice, for example from electronic health records. Most causal inference methods for longitudinal and time-to-event data have also focused on this simplified setting. However, recent work has been done by \cite{Ryalen:2019} to use MSMs based on additive hazard models in the continuous time setting, and by \cite{Seaman:2019}, who described the structural nested cumulative survival time models. It would be of interest to extend our simulation algorithm to the continuous time setting to enable comparisons involving these emerging methods.

	\begin{table}[!h]
		\caption{Cumulative coefficients at times $1,\ldots,5$ at times $1,\ldots,5$: true values, mean of the estimates (and empirical SE) obtained using MSM-IPTW from 1000 simulations, and bias in the estimates (and Monte Carlo SE) obtained using MSM-IPTW.}
		\label{tab:cumcoef}
		\centering
		{\small
			\begin{tabular}{rrrr}
				\hline
				&&\multicolumn{2}{c}{MSM-IPTW}\\
				\cline{3-4}
				Time&True value&Mean estimate (Empirical SE)&Bias (Monte Carlo SE)\\
				\hline
				\multicolumn{4}{l}{Cumulative coefficient $\int_{0}^{t}\tilde{\alpha}_{0}(s)ds$}\\
				1 & 0.700 (0.009) & 0.699 (0.016) &  -0.001 (0.000) \\ 
				2 & 1.408 (0.016) & 1.407 (0.028) &  -0.000 (0.001) \\ 
				3 & 2.128 (0.026) & 2.129 (0.045) &  0.002 (0.001) \\ 
				4 & 2.863 (0.040) & 2.867 (0.070) &  0.003 (0.002) \\ 
				5 & 3.623 (0.058) & 3.630 (0.110) &  0.007 (0.003) \\ 
				&&&\\
				\multicolumn{4}{l}{Cumulative coefficient $\int_{0}^{t}\tilde{\alpha}_{A,0}(s)ds$}\\
				1 & -0.198 (0.010) & -0.199 (0.037) &  -0.001 (0.001) \\ 
				2 & -0.396 (0.017) & -0.397 (0.065) &  -0.000 (0.002) \\ 
				3 & -0.594 (0.023) & -0.592 (0.100) &  0.001 (0.003) \\ 
				4 & -0.790 (0.033) & -0.788 (0.150) &  0.002 (0.005) \\ 
				5 & -0.987 (0.042) & -0.968 (0.231) &  0.018 (0.007) \\ 
				&&&\\
				\multicolumn{4}{l}{Cumulative coefficient $\int_{0}^{t}\tilde{\alpha}_{A,1}(s)ds$ (equal to zero for $t\leq1$)}\\
				2 & -0.098 (0.013) & -0.102 (0.057)  &  -0.004 (0.002) \\ 
				3 & -0.195 (0.021) & -0.206 (0.096)  &  -0.011 (0.003)  \\ 
				4 & -0.291 (0.030) & -0.303 (0.155)  &  -0.013 (0.005)  \\ 
				5 & -0.386 (0.039) & -0.390 (0.245)&  -0.005 (0.008) \\ 
				&&&\\
				\multicolumn{4}{l}{Cumulative coefficient $\int_{0}^{t}\tilde{\alpha}_{A,2}(s)ds$ (equal to zero for $t\leq2$)}\\
				3 & -0.077 (0.017) & -0.076 (0.078)  &  0.001 (0.002) \\ 
				4 & -0.153 (0.027) & -0.153 (0.139)  &  0.000 (0.004)  \\ 
				5 & -0.228 (0.039) & -0.222 (0.232) &  0.006 (0.007)  \\ 
				&&&\\ 
				\multicolumn{4}{l}{Cumulative coefficient $\int_{0}^{t}\tilde{\alpha}_{A,3}(s)ds$ (equal to zero for $t\leq3$)}\\
				4 & -0.060 (0.021) & -0.061 (0.115)  &  -0.001 (0.004) \\ 
				5 & -0.121 (0.035) & -0.128 (0.211)  &  -0.007 (0.007) \\  
				&&&\\ 
				\multicolumn{4}{l}{Cumulative coefficient $\int_{0}^{t}\tilde{\alpha}_{A,4}(s)ds$ (equal to zero for $t\leq4$)}\\
				5 & -0.047 (0.028) & -0.039 (0.176) & 0.008 (0.006) \\ 
				\hline
		\end{tabular}}
	\end{table}
	
	\begin{table}[!h]
		\caption{Survival probabilities for the treatment regimes `never treated' and `always treated' at times $1,\ldots,5$: true values, mean of the estimates (and empirical SE) obtained using MSM-IPTW from 1000 simulations, and bias in the estimates (and Monte Carlo SE) obtained using MSM-IPTW.}
		\label{tab:surv}
		\centering
		{\small
			\begin{tabular}{rrrr}
				\hline
				&&\multicolumn{2}{c}{MSM-IPTW}\\
				\cline{3-4}
				Time&True value&Mean estimate (Empirical SE)&Bias (Monte Carlo SE)\\
				\hline
				\multicolumn{4}{l}{Never treated: $\Pr(T^{\underline{a}_0=0}>t)$}\\
				1 & 0.497 & 0.497 (0.008) & 0.000 (0.000) \\ 
				2 & 0.245 & 0.245 (0.007) & 0.000 (0.000) \\ 
				3 & 0.119 & 0.119 (0.005) & -0.000 (0.000) \\ 
				4 & 0.057 & 0.057 (0.004) & -0.000 (0.000) \\ 
				5 & 0.027 & 0.027 (0.003) & -0.000 (0.000) \\  
				&&&\\
				\multicolumn{4}{l}{Always treated: $\Pr(T^{\underline{a}_0=1}>t)$}\\
				1 & 0.606 & 0.607 (0.021) & 0.001 (0.001) \\ 
				2 & 0.401 & 0.404 (0.031) & 0.003 (0.001) \\ 
				3 & 0.283 & 0.288 (0.040) & 0.005 (0.001) \\ 
				4 & 0.208 & 0.216 (0.051) & 0.007 (0.002) \\ 
				5 & 0.157 & 0.165 (0.066) & 0.009 (0.002) \\ 
				\hline
		\end{tabular}}
	\end{table}
	
	\clearpage
	
	\bibliographystyle{biom}
	\bibliography{references}

\textbf{Acknowledgements}

RHK is funded by a UK Research \& Innovation Future Leaders Fellowship (MR/S017968/1), SRS by MRC programme grant MC\_UU\_00002/10, and JMG by the Research Council of Norway (Grant No. 273674).

\pagebreak

\setcounter{section}{0}
\setcounter{equation}{0}
\setcounter{figure}{0}
\setcounter{table}{0}
\setcounter{page}{1}

\renewcommand{\thesection}{A\arabic{section}}
\renewcommand{\theequation}{A\arabic{equation}}
\renewcommand{\thefigure}{A\arabic{figure}}

\makeatletter
\makeatother

\begin{center}
{\LARGE Simulating longitudinal data from marginal structural models using the additive hazard model \\     \vspace{0.2cm} Supplementary materials}
    \vskip 1em%

\large{Ruth H. Keogh$^1$, Shaun R. Seaman$^2$, Jon Michael Gran$^3$, Stijn Vansteelandt$^{1,4}$}
	\vskip 1em%
{\small$^1$Department of Medical Statistics, London School of Hygiene \& Tropical Medicine, Keppel Street, London, WC1E 7HT, UK\\
$^2$MRC Biostatistics Unit, University of Cambridge, Institute of Public Health, Forvie Site, Robinson Way, Cambridge CB2 0SR, UK\\
$^3$Oslo Centre for Biostatistics and Epidemiology, Department of Biostatistics, Institute of Basic Medical Sciences, University of Oslo, P.O. Box 1122 Blindern, 0317 Oslo, Norway\\
$^4$Department of Applied Mathematics, Computer Science and Statistics, Ghent University, 9000 Ghent, Belgium}

\end{center}

\section{Inverse probability of treatment weights (IPTW)}
\label{sec:intro}

To estimate MSMs using IPTW, the weight at time $t$ for individual $i$ is the inverse of their probability of their observed treatment pattern up time time $t$ given their time-dependent covariate history \citep{Daniel:2013,Cole:2008}
\begin{equation}
W_i(t)=\prod_{k=0}^{\lfloor t \rfloor}\frac{1}{\Pr(A_k=A_{k,i}|\bar{L}_{k,i},\bar{A}_{k-1,i},T\geq k)}
\end{equation}
Some individuals can have very large weights, which can results in the parameters of the MSM being estimated very imprecisely, and therefore stabilized weights are typically used. The stabilized weight for individual $i$ is:
\begin{equation}
SW_i(t)=\prod_{k=0}^{\lfloor t \rfloor}\frac{\Pr(A_k=A_{k,i}|\bar{A}_{k-1,i},T\geq k)}{\Pr(A_k=A_{k,i}|\bar{L}_{k,i},\bar{A}_{k-1,i},T\geq k)}
\label{eq:stab.wt1}
\end{equation}

The MSMs in equations (3) and (4) of the main text are marginal over the distribution of the characteristics of the population at time $0$. It is also common to condition on baseline characteristics $L_0$, in which case the MSMs are of the form $\lambda_{T^{\underline{a}_{0}}}(t|L_0)=\lambda_{0}(t)\exp\lcb g(\bar{a}_{\lfloor t \rfloor},L_0;\beta)\rcb$ and $\lambda_{T^{\underline{a}_{0}}}(t|L_0)=\alpha_{0}(t)+ g(\bar{a}_{\lfloor t \rfloor},L_0;\alpha(t))$. The contributions of $L_0$ may be through main effects only, or there may be interactions between $L_0$ and $\bar{a}_{\lfloor t \rfloor}$. When the MSM is conditional on $L_0$, the numerator in the stabilized weights may also condition on $L_0$, and vice-versa:
\begin{equation}
SW_i(t)=\prod_{k=0}^{\lfloor t \rfloor}\frac{\Pr(A_k=A_{k,i}|\bar{A}_{k-1,i},L_{0,i},T\geq k)}{\Pr(A_k=A_{k,i}|\bar{L}_{k,i},\bar{A}_{k-1,i},T\geq k)}
\end{equation}

\section{MSMs using conditional additive hazard models: additional results}
\label{sec:details}

In this section we use the general results from Section 4.2 to derive the form of the MSM $\lambda_{\Tc}(t)$ when the conditional additive hazard is of the form 
\begin{equation}
\lambda(t|\bar A_{\lfloor t \rfloor},\bar L_{\lfloor t \rfloor},U)=\alpha_0(t)+\sum_{j=0}^{\lfloor t \rfloor} \alpha_{Aj}(t) A_{\lfloor t \rfloor-j}+\sum_{j=0}^{\lfloor t \rfloor} \alpha_{Lj}(t) L_{\lfloor t \rfloor-j}+\alpha_U(t) U
\label{eq:cond.haz}
\end{equation}
and when the covariates are normally and conditionally normally distributed as follows
$$
U\sim N(\nu,\phi^2)
$$
$$
L_0|U\sim N(\theta_{00} +\theta_{0U} U,\sigma^2_0)
$$
$$
L_1|A_0=a_0,L_0,U,T\geq 1\sim N(\theta_{10}+\theta_{1A}a_0+\theta_{1L}L_0 +\theta_{1U} U,\sigma^2_1)
$$
We use the following notation for the cumulative coefficients of the conditional additive hazard model
$$
\aUt=\int_{0}^{t}\alpha_U(s)ds,
$$
$$
\aLo=\int_{0}^{t}\alpha_{L0}(s)ds,\quad \aaLo=\int_{1}^{t}\alpha_{L0}(s)ds
$$
$$
\aLl=\int_{0}^{t}\alpha_{L1}(s)ds
$$
The results given below use the general results that for $X\sim N(\mu,\sigma^2)$
\begin{equation}
E\left\{\exp(-Xw)\right\}=\exp\left(-\mu w+\sigma^2w^2/2\right)
\label{eq:norm1}
\end{equation}
\begin{equation}
E\left\{X\exp(-Xw)\right\}=-\frac{d}{dw}E\left\{\exp(-Xw)\right\}=(\mu-\sigma^2 w)\exp\left(-\mu w+\sigma^2w^2/2\right)
\label{eq:norm2}
\end{equation} 

For $0<t<1$ the conditional hazard in (\ref{eq:cond.haz}) is $\lambda(t|A_{0},L_{0},U)=\alpha_0(t)+\alpha_{A0}(t) A_{0}+\alpha_{L0}(t) L_{0}+\alpha_U(t) U$. Using the result in (15) in the main text, the form of $\lambda_{\Tc}(t)$ for $0<t<1$ is
{\small
\begin{equation}
\begin{split}
\lambda_{\Tc}(t)&=\alpha_0(t)+\alpha_{A0}(t) a_0+\frac{E_{L_0,U}\left\{ \left(\alpha_{L0}(t) L_0+\alpha_U(t) U\right)\exp\left(-\aLo L_0-\aUt U\right)\right\}}{E_{L_0,U}\left\{ \exp\lb-\aLo L_0-\aUt U\rb\right\}}\\
&=\alpha_0(t)+\alpha_{A0}(t) a_0\\
&\qquad+\frac{E_{U}\left\{\exp\left(-\aUt U\right)\left[\alpha_{L0}(t)E_{L_0|U}\left\{ L_0\exp\left(-\aLo L_0\right)\right\}+\alpha_U(t) UE_{L_0|U}\left\{\exp\left(-\aLo L_0 \right)\right\}\right]\right\}}{E_{U}\left[\exp\left(-\aUt U\right) E_{L_0|U}\left\{\exp\left(-\aLo L_0 \right)\right\} \right]}
\end{split}
\label{eq:msm.1}
\end{equation}}

We let
$$
C=\exp\lb-\theta_{00} \aLo+\sigma^2_0\aLo^2/2\rb
$$
$$
D=\exp\lcb-\nu(\theta_{0U}\aLo+\aUt)+\phi^2(\theta_{0U}\aLo+\aUt)^2/2\rcb
$$
Under the assumed normal distributions for $U$ and $L_0|U$ and using the results in (\ref{eq:norm1}) and (\ref{eq:norm2}) it can be shown that
{\small
\begin{eqnarray}
\nonumber E_{L_0|U}\lcb\exp\lb-\aLo L_0 \rb \rcb&=&C\exp\lb-\theta_{0U} \aLo U\rb\\
\nonumber E_{L_0|U}\lcb L_0\exp\lb-\aLo L_0\rb\rcb&=&\lb\theta_{00} +\theta_{0U} U-\sigma^2_0 \aLo\rb C\exp\lb-\theta_{0U} \aLo U\rb\\
\nonumber E_{U}\lsb\exp\lb-\aUt U\rb E_{L_0|U}\lcb \exp\lb-\aLo L_0\rb\rcb\rsb&=&CD\\
\nonumber E_{U}\lsb U\exp\lb-\aUt U\rb E_{L_0|U}\lcb \exp\lb-\aLo L_0\rb\rcb\rsb&=&CD\left(\nu-\phi^2\aUt-\phi^2\theta_{0U}\aLo \right)\\
\nonumber E_{U}\lsb \exp\lb-\aUt U\rb E_{L_0|U}\lcb L_0 \exp\lb-\aLo L_0\rb\rcb\rsb&=&CD\lcb\theta_{00}-\sigma^2_0\aLo+\theta_{0U}\lb\nu-\phi^2\aUt-\phi^2\theta_{0U}\aLo\rb\rcb
\end{eqnarray}}
It follows that $\lambda_{\Tc}(t)$ for $0<t<1$ in (\ref{eq:msm.1}) can be written 
\begin{equation}
\lambda_{\Tc}(t)=\tilde{\alpha}_0(t)+\alpha_{A0}(t) a_0
\end{equation}
where
{\small
\begin{equation}
\tilde{\alpha}_0(t)=\alpha_0(t)+\alpha_{L0}(t)\lb\theta_{00}-\sigma^2_0\aLo\rb+\lb\alpha_{L0}(t)\theta_{0U}+\alpha_U(t)\rb\lb\nu-\phi^2\aUt-\phi^2\theta_{0U}\aLo\rb
\end{equation}}

For $1\leq t<2$ the conditional hazard in (\ref{eq:cond.haz}) is $\lambda(t|\bar A_{1},\bar L_{1},U)=\alpha_0(t)+\alpha_{A0}(t) A_{1}+\alpha_{A1}(t) A_{0}+\alpha_{L0}(t) L_{1}+\alpha_{L1}(t) L_{0}+\alpha_U(t) U$. Using the result in (16) in the main text, the form of $\lambda_{\Tc}(t)$ for $1\leq t<2$ is
\begin{equation}
{\footnotesize
	\begin{split}
	\lambda&_{\Tc}(t)= \alpha_0(t)+\alpha_{A0}(t) a_1+\alpha_{A1}(t) a_0+\\
	&\frac{E_{L_0,U}\lsb E_{L_1|A_0=a_0,L_0,U,T\geq 1}\lcb \lb\alpha_{L0}(t)L_1+\alpha_{L1}(t)L_0+\alpha_U(t)U\rb \exp\lb-\aaLo L_1 -\aLl L_0-\aUt U\rb\rcb\rsb}{E_{L_0,U}\lsb E_{L_1|A_0=a_0,L_0,U,T\geq 1}\lcb\exp\lb-\aaLo L_1-\aLl L_0-\aUt U\rb\rcb\rsb}\\
&= \alpha_0(t)+\alpha_{A0}(t)a_1+\alpha_{A1}(t) a_0+\\
	&\frac{E_{U}\left\{\exp\lb-\aUt U\rb E_{L_0|U}\lsb\exp\lb-\aLl L_0\rb E_{L_1|A_0=a_0,L_0,U,T\geq 1}\lcb \lb\alpha_{L0}(t)L_1+\alpha_{L1}(t)L_0+\alpha_U(t)U\rb \exp\lb-\aaLo L_1 \rb\rcb\rsb\right\}}{E_{U}\lcb \exp\lb-\aUt U\rb E_{L_0|U}\lsb\exp\lb-\aLl L_0\rb E_{L_1|A_0=a_0,L_0,U,T\geq 1}\lcb\exp\lb-\aaLo L_1 \rb\rcb\rsb\rcb}
	\end{split}}
\label{eq:msm2}
\end{equation}

We let
\begin{equation}
\nonumber \begin{split}
F&=\exp\lcb-(\theta_{10}+\theta_{1A}a_0)\aaLo+\sigma^2_1\aaLo^2/2\rcb\\
G&=\exp\lcb-\theta_{00}\lb\aLl+\theta_{1L}\aaLo\rb+\sigma^2_0\lb\aLl+\theta_{1L}\aaLo\rb^2/2\rcb\\
H&=\exp\lcb-\nu\lb\aUt+\theta_{1U}\aaLo+\theta_{0U}\aLl+\theta_{0U}\theta_{1L}\aaLo\rb\right.\\
&\left.\qquad\qquad+\phi^2\lb\aUt+\theta_{1U}\aaLo+\theta_{0U}\aLl+\theta_{0U}\theta_{1L}\aaLo\rb^2/2\rcb\\
J&=\alpha_{L0}(t)\lb\theta_{10}-\sigma^2_1\aaLo\rb+\lb\alpha_{L1}(t)+\alpha_{L0}(t)\theta_{1L}\rb\{\theta_{00}-\sigma^2_0\lb\theta_{1L}\aaLo+\aLl\rb\}\\
K&=\alpha_{U}(t)+\alpha_{L0}(t)\theta_{1U}+\theta_{0U}\lb\alpha_{L1}(t)+\alpha_{L0}(t)\theta_{1L}\rb
\end{split}
\end{equation}

Under the assumed normal distributions for $U$, $L_0|U$, $L_1|A_0=a_0,L_0,U,T\geq 1$ and using the results in (\ref{eq:norm1}) and (\ref{eq:norm2}), the term in the denominator of the ratio of expectations in the third term of (\ref{eq:msm2}) can be derived sequentially as follows: 
\begin{equation}
{\small
\nonumber \begin{split}
E_{L_1|A_0=a_0,L_0,U,T\geq 1}\lcb\exp\lb-\aaLo L_1 \rb\rcb&=F\exp\lb-\theta_{1L} \aaLo L_0-\theta_{1U}\aaLo U\rb\\
E_{L_0|U}\lsb\exp\lb-\aLl L_0\rb E_{L_1|A_0=a_0,L_0,U,T\geq 1}\lcb\exp\lb-\aaLo L_1\rb\rcb\rsb&=FG\times\\
&\hspace{-2cm}  \exp\lcb-\lb\theta_{1U}\aaLo+\theta_{0U}\theta_{1L}\aaLo+\theta_{0U}\aLl\rb U\rcb
\end{split}}
\end{equation}
\begin{equation}
\nonumber 
E_{U}\lcb\exp\lb-\aUt U\rb E_{L_0|U}\lsb \exp\lb-\aLl L_0\rb E_{L_1|A_0=a_0,L_0,U,T\geq 1}\lcb \exp\lb-\aaLo L_1 \rb\rcb\rsb\rcb=FGH
\end{equation}
Similarly, the terms in the numerator of the ratio of expectations in the third term of (\ref{eq:msm2}) can be derived sequentially as follows: 
\begin{equation}
\nonumber \begin{split}
E_{L_1|A_0=a_0,L_0,U,T\geq 1}\lcb \lb\alpha_{L0}(t)L_1+\alpha_{L1}(t)L_0+\alpha_U(t)U\rb \exp\lb-\aaLo L_1\rb\rcb&=\\
&\hspace{-6cm}\exp\lb-\theta_{1L} \aaLo L_0-\theta_{1U}\aaLo U\rb\times\\
 F\lcb\alpha_{L0}(t)\lb\theta_{10}+\theta_{1A}a_0-\sigma^2_1\aaLo\rb+\lb\alpha_{L1}(t)+\alpha_{L0}(t)\theta_{1L}\rb L_0+\lb\alpha_{U}(t)+\alpha_{L0}(t)\theta_{1U}\rb U\rcb
 \end{split}
 \end{equation}
 \begin{equation}
 \nonumber \begin{split}
E_{L_0|U}\lsb \exp\lb-\aLl L_0\rb E_{L_1|A_0=a_0,L_0,U,T\geq 1}\lcb \lb\alpha_{L0}(t)L_1+\alpha_{L1}(t)L_0+\alpha_U(t)U\rb \exp\lb-\aaLo L_1 \rb\rcb\rsb=\\
F G\lb J+\alpha_{L0}(t)\theta_{1A}a_0+KU\rb\exp\lcb-\lb\theta_{0U}\theta_{1L}\aaLo+\theta_{0U}\aLl+\theta_{1U}\aaLo\rb U\rcb
\end{split}
\end{equation}
{\footnotesize
	\begin{equation}
	\nonumber \begin{split}
E_{U}\lcb \exp\lb-\aUt U\rb E_{L_0|U}\lsb \exp\lb-\aLl L_0\rb E_{L_1|A_0=a_0,L_0,U,T\geq 1}\lcb \lb\alpha_{L0}(t)L_1+\alpha_{L1}(t)L_0+\alpha_U(t)U\rb \exp\lb-\aaLo L_1 \rb\rcb\rsb\rcb=\\
F GH\lsb J+\alpha_{L0}(t)\theta_{1A}a_0+K\lcb\nu-\phi^2\lb \aUt+\theta_{1U}\aaLo+\theta_{0U}\aLl+\theta_{0U}\theta_{1L}\aaLo\rb \rcb\rsb
\end{split}
\end{equation}}
It can be shown using the above results that 
\begin{equation}
	\lambda_{\Tc}(t)= \tilde{\alpha}_0(t)+\alpha_{A0}(t)a_1+\tilde\alpha_{A1}(t)a_0
\end{equation}
where
$$
\tilde{\alpha}_0(t)=\alpha_0(t)+K\lcb\nu-\phi^2\lb \aUt+\theta_{1U}\aaLo+\theta_{0U}\aLl+\theta_{0U}\theta_{1L}\aaLo\rb \rcb+J
$$
and 
$$
\tilde{\alpha}_{A1}(t)=\alpha_{A1}(t)+\alpha_{L0}(t)\theta_{1A}
$$

We have therefore derived the form of the MSM $\lambda_{\Tc}(t)$ for $0<t<1$ and $1\leq t<2$ when the conditional hazard is of the form in (\ref{eq:cond.haz}) and when the covariates are normally and conditionally normally distributed. As shown in more general results in Section 4.2 of the main text, the MSMs have an additive form. However, the above results show that the formulae for the coefficients in the MSM take quite a complicated form even in this relatively simple setting. The expressions would become further complicated if there were multiple time-dependent covariates $L$ and when the conditional distributions for the covariates given the past were not normal, in which case there will not in general exist closed form expressions for the coefficients of the MSM. In Section 6.2 of the main text we outline a simulation-based procedure for obtaining the true values of the coefficients in the MSM. 

\section{Incorporating interactions}
\label{sec:interaction}

In Section 4.2 of the main text, we considered the conditional additive hazard model given in equation (14). Suppose instead that there was also an interaction between $\bar A_t$ and $\bar L_t$:
\begin{equation}
\lambda(t|\bar A_{\lfloor t \rfloor},\bar L_{\lfloor t \rfloor},U)=\alpha_0(t)+\alpha_A^{\top}(t)v(\bar A_t)+\alpha_L^{\top}(t)w(\bar L_t)+\alpha_{AL}^{\top}(t)q(\bar A_t,\bar L_t)+\alpha_U(t)U
\end{equation}
where $q(\bar A_{\lfloor t \rfloor},\bar L_{\lfloor t \rfloor})$ denotes a vector values function of interactions between $\bar A_t$ and $\bar L_t$. Following the same workings as in Section 4.2 of the main text, it can be shown that for $0<t<1$
	\begin{equation}
	\lambda_{\Tc}(t)= \alpha_0(t)+\alpha_A^{\top}(t)v(a_0)+\frac{E_{L_0,U}\lcb \left(\alpha_L^{\top}(t)w(L_0)+\alpha_{AL}^{\top}(t)q(a_0,L_0)+\alpha_U(t)U\right)r_0(t)\rcb}{E_{L_0,U}\lcb r_0(t)\rcb}
	\end{equation}
where $r_0(t)=\exp\lb-\int_{0}^{t}(\alpha_L^{\top}(s)w(L_0)+\alpha_{AL}^{\top}(s)q(a_0,L_0)+\alpha_U(s)U)ds\rb$. 

For $1\leq t <2$ we have
\begin{equation}
{\small
	\begin{split}
	\lambda&_{\Tc}(t)= \alpha_0(t)+\alpha_A^{\top}(t)v(\bar{a}_1)+\\
	&\qquad\qquad\frac{E_{L_0,U}\lsb E_{L_1|A_0=a_0,L_0,U,T\geq 1}\lcb \lb\alpha_L^{\top}(t)w(\bar L_1)+\alpha_{AL}^{\top}(t)q(\bar a_1,\bar L_1)+\alpha_U(t)U\rb r_1(t)\rcb\rsb}{E_{L_0,U}\lsb E_{L_1|A_0=a_0,L_0,U,T\geq 1}\lcb r_1(t)\rcb\rsb}
	\end{split}}
\end{equation}
where $r_1(t)=r_0(t)\exp\lcb-\int_{1}^{t}\lb\alpha_L^{\top}(s)f(\bar{L}_1)+\alpha_{AL}^{\top}(s)q(\bar a_1,\bar L_1)+\alpha_U(s)U\rb ds\rcb$.

For $1\leq t<2$ the intercept and the coefficients for $a_0$ and $a_1$ in the MSM are different from those in the conditional model. The MSM also involves an interaction between $a_0$ and $a_1$ even if there is no interaction between $a_0$ and $a_1$ in the conditional hazard model.

\section{Simulation algorithm: extensions}
\label{sec:sim.extra}

In section 5 of the main text we described a simulation algorithm for longitudinal and time-to-event data, using a conditional additive hazard model of the form $\lambda(t|\bar A_{\lfloor t \rfloor},\bar L_{\lfloor t \rfloor},U)=\alpha_0+\alpha_A A_{\lfloor t \rfloor}+\alpha_L L_{\lfloor t \rfloor}+\alpha_U U$. The algorithm can be extended to accommodate a more general form for the conditional hazard including time-varying coefficients: $\lambda(t|\bar A_{\lfloor t \rfloor},\bar L_{\lfloor t \rfloor},U)=\alpha_0(t)+\alpha_A^{\top}(t)v(\bar A_{\lfloor t \rfloor})+\alpha_L^{\top}(t)w(\bar L_{\lfloor t \rfloor})+\alpha_U(t)U$. For the simulation the investigator needs to specify the functional forms for the coefficients. One way to simulate data in this more general setting is by generating event times using a piecewise exponential distribution, as we outline below. Further extensions to include additional terms such as interaction terms follow directly. 

A general form for the simulation algorithm is as follows:
\begin{enumerate}
	\item Generate the individual frailty term $U$.
	\item Generate $L_0$ conditional on $U$.
	\item Generate $A_0$ from a Bernoulli distribution conditional on $L_0$.
	\item The conditional hazard is $\lambda(t|\bar A_{\lfloor t \rfloor},\bar L_{\lfloor t \rfloor},U)=\alpha_0(t)+\alpha_A^{\top}(t)v(\bar A_{\lfloor t \rfloor})+\alpha_L^{\top}(t)w(\bar L_{\lfloor t \rfloor})+\alpha_U(t)U$. Event times are generated in the period $0<t< 1$ using a piecewise exponential distribution on a grid from 0 to 1 in increments of length 0.1 (this could be made smaller or larger). The procedure is as follows. First generate $V\sim \mathrm{Uniform}(0,1)$ and calculate $T^*=-\log(V)/\lambda(0|A_0,L_0,U)$. If $T^*<0.1$ the event time is set to be $T=T^*$. If $T^*>0.1$, then for $w=0.1,0.2,\ldots,0.9$:	
		\begin{itemize}
			\item[(i)] Generate $v\sim \mathrm{Uniform}(0,1)$ and calculate $T^*=-\log(V)/\lambda(w|A_0,L_0,U)$.
			\item[(ii)] If $T^*<0.1$ the event time is set to be $T=w+T^*$. 
			\item[(iii)] If $T^*>0.1$ move to the next value of $w$ and return to (i).
			\item[(iv)] When $w=0.9$, if $T^*>0.1$ move to step 5. 
		\end{itemize}	
	For individuals who remain at risk of the event at visit time $k=1$:
	\item Generate $L_k$ conditional on $\bar A_{k-1}, \bar L_{k-1},U,T\geq k$. 
	\item Generate $A_k$ from a Bernoulli distribution conditional on $\bar A_{k-1}, \bar L_{k},T\geq k$.
	\item Generate event times in the period $k\leq t< k+1$ using a piecewise exponential distribution on a grid from $k$ to $k+1$ in increments of length 0.1. First generate $V\sim \mathrm{Uniform}(0,1)$ and calculate $T^*=-\log(V)/\lambda(k|\bar{A}_1,\bar{L}_1,U)$. If $T^*<0.1$ the event time is set to be $T=T^*$. If $T^*>0.1$, then for $w=k+0.1,k+0.2,\ldots,k+0.9$:
		\begin{itemize}
			\item[(i)] Generate $V\sim \mathrm{Uniform}(0,1)$ and calculate $T^*=-\log(V)/\lambda(w|\bar{A}_k,\bar{L}_k,U)$.
			\item[(ii)] If $T^*<0.1$ the event time is set to be $T=w+T^*$. 
			\item[(iii)] If $T^*>0.1$ move to the next value of $w$ and return to (i).
			\item[(iv)] When $w=k+9$, if $T^*>0.1$ the individual remains at risk of the event at time $k+1$.
		\end{itemize}	
	\item Repeat steps 5-7 for $k=2,3,4$. Individuals who do not have an event time generated in the period $0<t<5$ are administratively censored at time 5. 
\end{enumerate}

%\bibliography{references_supp_arxiv}
	
\end{document}